\documentclass[a4paper]{article}
\usepackage[english]{babel}
\usepackage[utf8x]{inputenc}
\usepackage[T1]{fontenc}

\usepackage[letterpaper,top=1.5cm,bottom=2cm,left=1.5cm,right=1.5cm,marginparwidth=1.75cm]{geometry}

\usepackage{amsmath}
\usepackage{graphicx}
\usepackage[colorinlistoftodos]{todonotes}
\usepackage[colorlinks=true, allcolors=blue]{hyperref}
\usepackage{siunitx}
\usepackage{authblk}
\usepackage{xcolor}
\usepackage{nameref}
\usepackage{afterpage}

\usepackage[sort&compress,numbers]{natbib}
\usepackage{setspace}
\title{PyBioNetFit and the Biological Property Specification Language}
\author{Eshan D. Mitra, Ryan Suderman, Joshua Colvin, Alexander Ionkov, Andrew Hu, Herbert M. Sauro, Richard G. Posner, and William S. Hlavacek}

\date{March 2019}

\usepackage{natbib}
\usepackage{graphicx}

\begin{document}

\maketitle

\section*{Abstract}

In systems biology modeling, important steps include model parameterization, uncertainty quantification, and evaluation of agreement with experimental observations. To help modelers perform these steps, we developed the software PyBioNetFit. PyBioNetFit is designed for parameterization, and also supports uncertainty quantification, checking models against known system properties, and solving design problems. PyBioNetFit introduces the Biological Property Specification Language (BPSL) for the formal declaration of system properties. BPSL allows qualitative data to be used alone or in combination with quantitative data for parameterization model checking, and design. PyBioNetFit performs parameterization with parallelized metaheuristic optimization algorithms (differential evolution, particle swarm optimization, scatter search) that work directly with existing model definition standards:  BioNetGen Language (BNGL) and Systems Biology Markup Language (SBML). We demonstrate PyBioNetFit’s capabilities by solving 31 example problems, including the challenging problem of parameterizing a model of cell cycle control in yeast. We benchmark PyBioNetFit's parallelization efficiency on computer clusters, using up to 288 cores. Finally, we demonstrate the model checking and design applications of PyBioNetFit and BPSL by analyzing a model of therapeutic interventions in autophagy signaling.

\ 

\textbf{Keywords:} Categorical data / Formal verification / Parallel computing / Parameter identification / Rule-based modeling

\section{Introduction}




An important step in the development of a mathematical model for a biological system is using experimental data to identify model parameters. In a conventional approach, the experimental data of most utility are quantitative time-courses and/or dose-response curves. Parameters are adjusted to minimize the difference between the model outputs and the experimental data (as measured, for example, by a residual sum of squares function). 

In some cases, there are straightforward solutions for parameter identification. For models that can be expressed as a small system of differential equations (e.g., tens of equations, as in refs. \cite{Raue2013, Hass2018}), general-purpose software tools are available such as Data2Dynamics \cite{Raue2015} and COPASI \cite{Hoops2006} that can apply, for example, gradient-based methods to find an optimal parameter set. However, not all biological models fall into this category. When current software tools are inadequate, modelers must resort to either problem-specific code or manual adjustment of parameters. Both of these approaches are tedious from the perspective of the modeler, and also present challenges for reproducibility of the modeling work \cite{Medley2016,Waltemath2016}. Therefore, there is strong motivation to expand the scope of problems that can be solved using general-purpose software compatible with standard model definition formats. 

We developed the software PyBioNetFit to solve three major classes of parameterization problems for which current software solutions are limited. (i) Problems with exceptionally large systems of ODEs, where gradient-based methods are computationally expensive. Such problems often arise when using rule-based modeling. Rule-based modeling is the preferred approach for processes in which a combinatorial explosion in the number of possible chemical species makes it challenging to enumerate every possible chemical reaction \cite{Faeder2005,Chylek2013}. In a rule-based model, a concise set of rules can imply a much larger system of ODEs (hundreds to thousands of equations).  (ii) Problems featuring models that are simulated stochastically. This class of problems includes rule-based models in which the implied ODE system is so large that it cannot be derived from rules or numerically integrated efficiently \cite{Sneddon2011, Suderman2018}. In such cases, gradient calculations must use the finite difference method, which is inefficient \cite{Raue2013}. (iii) Problems including unconventional experimental data, in particular non-numerical \emph{qualitative data}. Such datasets are often collected by experimentalists, and have the potential to inform model parameterization \cite{Mitra2018a}, but currently are rarely used in practice. Notable exceptions are works of Tyson and coworkers \cite{Chen2000,Chen2004,Csikasz-Nagy2006,Oguz2013,Kraikivski2015} and Pargett et al. \cite{Pargett2013,Pargett2014}.

We solve problems (i) and (ii) by using parallelized metaheuristic optimization algorithms in place of gradient-based algorithms. Metaheuristics are a well-established class of optimization algorithms that do not rely on gradient information. Although they carry no guarantee for convergence to a global optimum, they are found to be effective in many use cases \cite{Gandomi2013}. Examples of metaheuristics include differential evolution \cite{Storn1997}, particle swarm optimization \cite{Eberhart1995}, and scatter search \cite{Glover2000}. Such algorithms often include some type of iterative randomized selection of candidate parameter sets, followed by evaluation of the selected parameter sets, which is used to direct the selection of parameters in future iterations to more favorable regions of parameter space. Many modern descriptions of metaheuristics allow for parallelized evaluation of parameter sets \cite{Penas2015a,Penas2017,Moraes2015}, which is valuable when each model simulation is computationally expensive. Although these algorithms are well-established, software designed for biological applications is limited. COPASI \cite{Hoops2006} and Data2Dynamics \cite{Raue2015} both include some metaheuristic algorithms, but these algorithms are not parallelized and have not been demonstrated on large ODE systems derived from rule-based models. The software BioNetFit \cite{Thomas2016} (called BioNetFit 1 in this report to distinguish it from the newly developed software) was an early effort to use a parallelized evolutionary algorithm to parameterize rule-based biological models. However, the BioNetFit 1 algorithm is inefficient in many cases, and in general, optimization algorithm performance is problem-dependent, and so a toolbox of methods is needed to enable a wide range of problems to be solved efficiently. PyBioNetFit was inspired by BioNetFit 1 but is an entirely new code base that includes multiple, robust metaheuristic algorithms. 

We solve problem (iii) by following our approach presented in ref. \cite{Mitra2018a} to parameterize models using both qualitative and quantitative data. In this approach, properties of interest are represented as one or more inequality constraints on the outputs of a model, enforced during some portion of a simulation. In some cases, a single qualitative observation, such as the viability of a particular mutant, implies several system properties (inequalities), such as upper bounds on several model outputs. After defining inequalities, we cast each inequality as a static penalty function \cite{Smith1997}, added to the objective function to be minimized. The result is a scalar-valued objective function with contributions from both qualitative and quantitative data; this function is minimized during fitting. This approach intentionally allows for the possibility that some of the inequality constraints may not be satisfied (because they arise from uncertain experimental data). 

To extend this approach for use in general-purpose software, we require a language to express arbitrary system properties of interest. In systems biology, there is no established means for formalizing system properties, although attempts have been made to do so with temporal logic \cite{Clarke2008,Heath2008,Kwiatkowska2008,David2012}, sometimes as part of model parameterization \cite{Liu2017,Hussain2015,Khalid2018}. However, there is a lack of software tools tailored for biological modeling that support property specification languages---most applications instead use problem-specific code. Additionally, there are few demonstrations of how the formalisms of temporal logic, originally developed for computer science applications \cite{Clarke1983}, can be applied to describe biologically interesting properties such as case-control comparisons. To address these deficiencies, we developed the Biological Property Specification Language (BPSL) as part of PyBioNetFit. BPSL is a domain-specific language for declaration of biological system properties, and allows such properties to be used as part of parameterization.

To complement its parameterization features, PyBioNetFit includes methods for uncertainty quantification of parameter estimates. Bayesian uncertainty quantification can be performed using Markov chain Monte Carlo (MCMC) with the Metropolis-Hastings algorithm (reviewed in \cite{Chib1995}) or parallel tempering (reviewed in \cite{Earl2005}). These methods start with an assumed prior probability distribution for each parameter, and aim to sample the multidimensional posterior probability distribution of the parameters given the data. The model can be simulated using sampled parameter sets to quantify the uncertainty of model predictions. Another uncertainty quantification method is bootstrapping \cite{Efron1994,Press2007}. Bootstrapping is a resampling procedure in which many replicates of fitting are performed, and each replicate uses a random sample of the experimental data available. The resulting list of best-fit parameter sets for the replicate problems is used to estimate confidence levels for best-fit parameter estimates.

Although PyBioNetFit and BPSL were designed primarily for model parameterization, BPSL also enables model checking and design. In model checking \cite{Clarke1999}, we seek to simply verify whether a model reproduces a set of known properties. Applications of formal model checking to biological processes have been considered, including for stochastic models \cite{Clarke2008,Heath2008,Kwiatkowska2008}. Much more often, model checking in biology is done informally as part of building a model. However, as models become more detailed, with an increasing number of known properties, a more formal and systematic system of model checking is useful: it can help in communicating what knowledge went into building the model, and for comparing the predictions of different models. Design represents a related application, analogous to the classical use of constrained optimization techniques. In a design problem in PyBioNetFit, we seek an intervention (a perturbation of a parameterized model) that brings about a desired set of BPSL-defined system behaviors; for example, choosing drug doses to up- or down-regulate activity of a target pathway.

All of the above features of PyBioNetFit are designed to be used in conjunction with existing model definition standards, avoiding the need for problem-specific code. PyBioNetFit natively supports models defined in BNGL \cite{Faeder2009}, a language for rule-based models, and SBML \cite{Hucka2003}, a language for more conventional models. For BNGL models, PyBioNetFit supports the ODE, SSA (stochastic simulation algorithm), and NFsim (network-free) \cite{Sneddon2011} simulators built into BioNetGen \cite{Harris2016,Faeder2009,Blinov2004}. For SBML models, PyBioNetFit uses the simulator libRoadRunner \cite{Somogyi2015}. PyBioNetFit has a modular design that makes it possible to add support for additional model standards and simulators in the future. Other model standards may be indirectly supported by converting to BNGL or SBML. For example, rule-based models defined in the Kappa language \cite{Danos2004, Sorokina2013} can be converted to BNGL using the software tool TRuML \cite{Suderman2017}.

To demonstrate the capabilities of PyBioNetFit, we solved a series of example optimization problems. We solved a total of 31 problems, 25 of which featured published, biologically relevant models \cite{Kozer2013, Monine2010, Chylek2014a, Harmon2017, Erickson2019, Faeder2003a, Romano2014, Blinov2006, Kocieniewski2012, Mitra2018a, Dunster2014, Boehm2014,Brannmark2010,Zheng2012a,Fey2015,Webb2011,Mukhopadhyay2013,Lee2003,Suderman2013,Kuhn2016,Hlavacek2018,Shirin2018,Oguz2013}. With four of these problems, we performed extensive benchmarking using different algorithms and different levels of parallelization. Not surprisingly, we find that the optimal algorithm depends on the fitting problem, which demonstrates the value of having a toolbox of several algorithms available. We then focus on a particularly challenging example problem: parameterizing the model of Tyson and co-workers for cell cycle control in yeast \cite{Chen2000,Chen2004,Csikasz-Nagy2006,Oguz2013,Kraikivski2015}. This model was originally parameterized by hand-tuning \cite{Chen2000,Chen2004} and later by automated optimization with problem-specific code \cite{Oguz2013, Mitra2018a}. Here we consider our most recent description of the problem \cite{Mitra2018a}, which has a 153-dimensional parameter space. We define the problem in BPSL, and solve it using the general-purpose functionality of PyBioNetFit. Thus, we demonstrate that PyBioNetFit can solve this general class of problem, that of using both qualitative and quantitative data to parameterize a biological model.

Finally, we considered a model describing drug intervention in autophagy signaling \cite{Shirin2018} to demonstrate the capabilities of PyBioNetFit and BPSL beyond model parameterization. We show that BPSL can be used to define a set of system properties, which can then be used in model checking. We also demonstrate how BPSL can be used to configure a design problem, finding a combination of drug doses to achieve a desired level of autophagy regulation. 

\section{Results}

\subsection{Workflow enabled by PyBioNetFit}
\label{sec:config}

\begin{figure}[tbp!]
\centering
\includegraphics{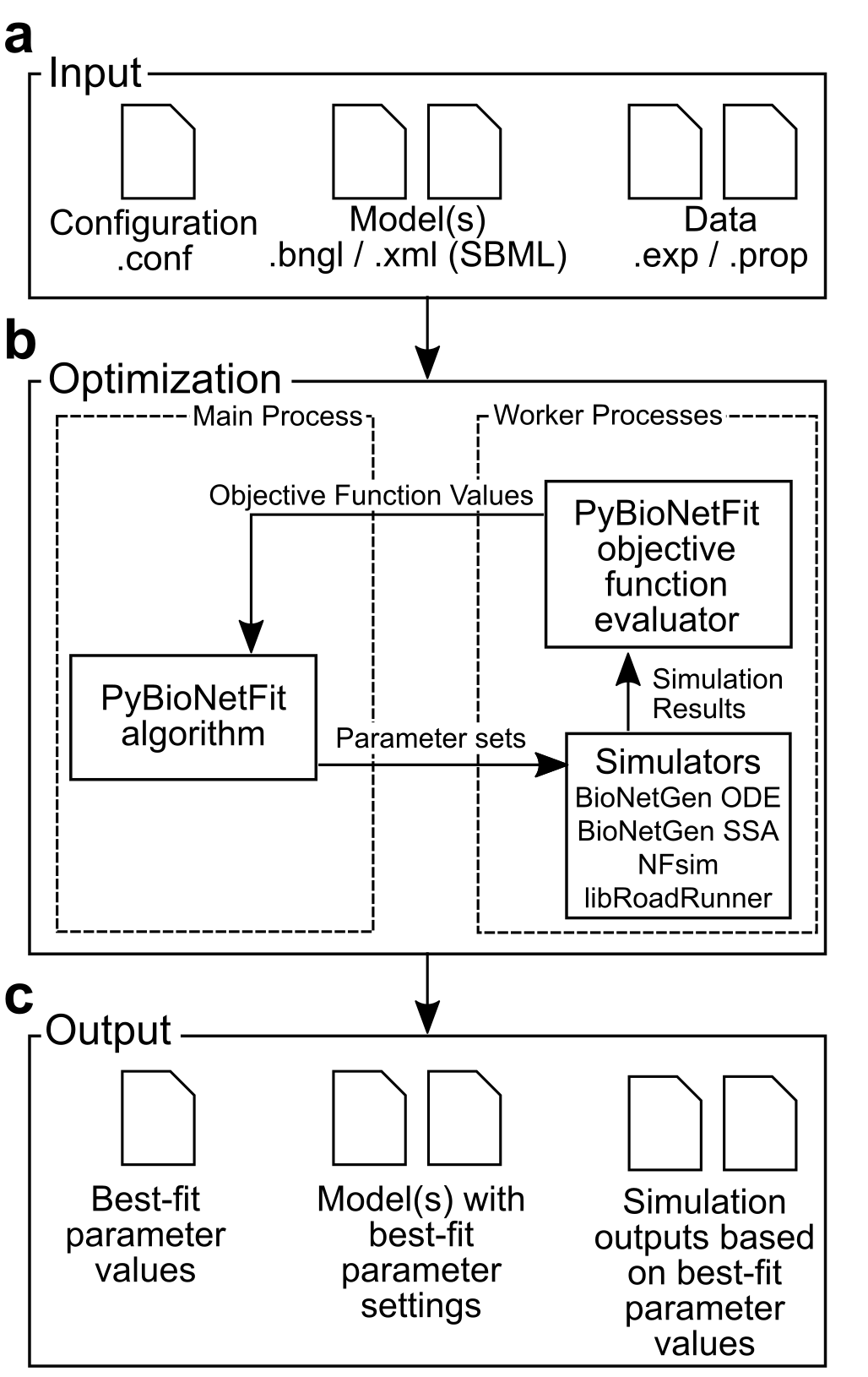}
\caption{Inputs, outputs, and operations of PyBioNetFit. (a) PyBioNetFit input files are a set of plain-text files: a CONF file specifying program settings, one or more model files in BNGL and/or SBML format, and one or more data files containing experimental data. EXP files contain quantitative data and PROP files contain qualitative data. Examples of these files are shown in Fig. \ref{fig:demob}. (b) When running PyBioNetFit, the user-selected optimization algorithm generates candidate parameter sets, which are passed to the appropriate simulator (for SBML models, libRoadRunner; for BNGL models, the simulator selected in the BNGL file). PyBioNetFit calculates the value of a user-selected objective function from the simulation results obtained for each trial parameter set, which is then used to inform future iterations of the algorithm. Each simulation and objective function evaluation is run in a separate worker process, which is run on a separate core of a multicore workstation or cluster if available. (c) PyBioNetFit output files include a text file reporting the best-fit parameter values, model files with the best-fit parameter settings, and output files resulting from simulating the models using the best-fit parameter values.}
\label{fig:demoa}
\end{figure}

The steps involved in using PyBioNetFit are illustrated in Figure \ref{fig:demoa}. PyBioNetFit is configured with a set of plain-text input files (Fig. \ref{fig:demoa}a). The input files must have particular filename extensions: .conf, .bngl, .xml, .exp, and .prop. We will refer to these as CONF files, BNGL files, etc. The files may be prepared in any standard text editor. 

\begin{figure}[tbp!]
\centering
\includegraphics{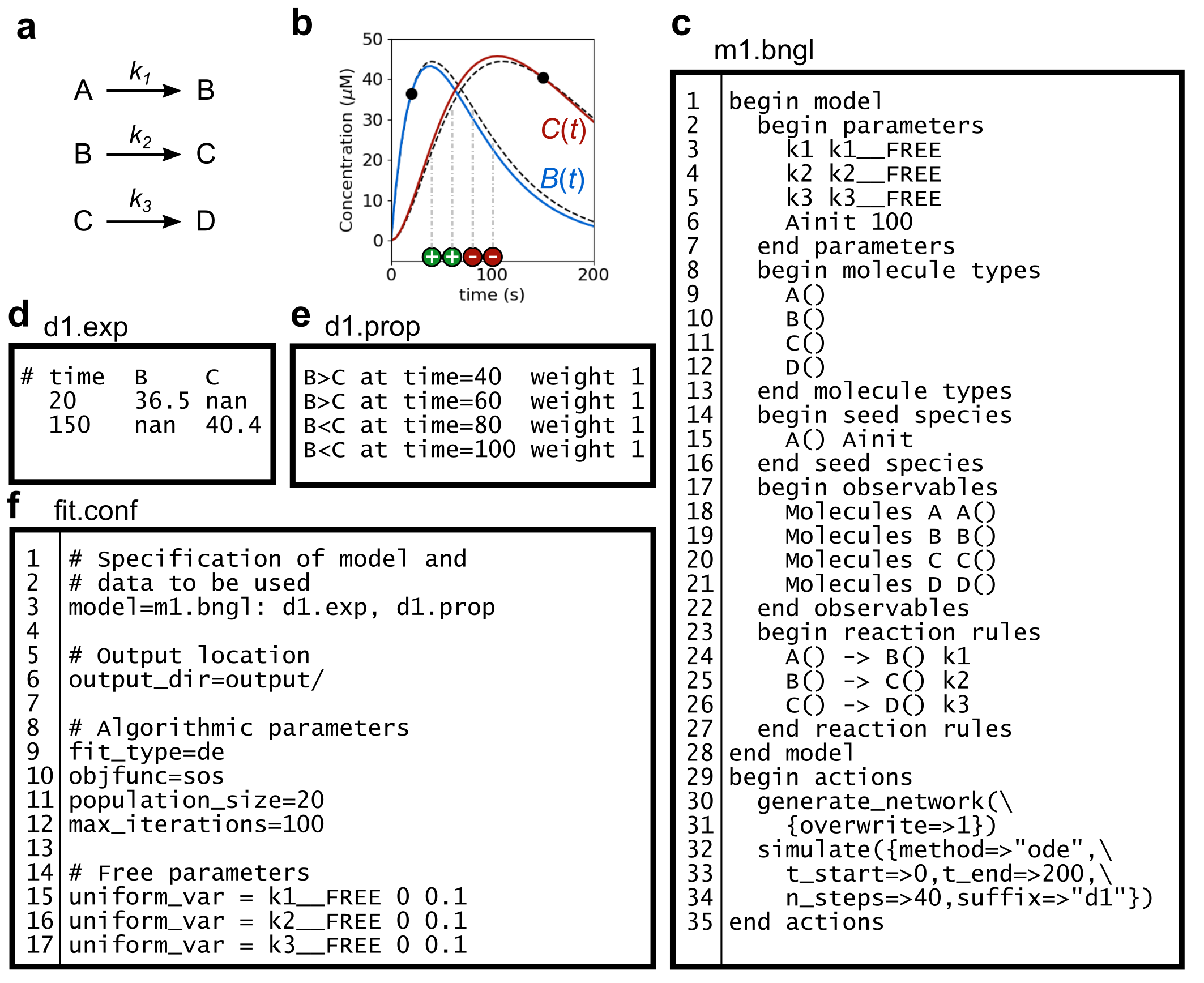}
\caption{A fitting problem configured to run in PyBioNetFit. (a) Reaction scheme of the model to be parameterized. The model is a coupled system of ODEs for the mass-action kinetics of the reactions shown here. The problem is to estimate values for the rate constants $k_1$, $k_2$, and $k_3$ (b) Time courses of concentrations of species B and C. Black broken curves give the ground truth. For fitting, two quantitative data points are available (black points), and four qualitative data points are available (colored circles). The qualitative data indicate whether the concentration of B or C is larger at a particular time. A plus sign indicates $B>C$ and a minus sign indicates $B<C$. Colored curves show the quality of fit after running PyBioNetFit on the input files listed in panels c-f. (c) Implementation in BNGL of a model for the reaction scheme in (a). Note that the parameters to be tuned by PyBioNetFit are named with the suffix \texttt{\_\_FREE} (lines 3-5). The parameter \texttt{Ainit}, which is to be held fixed, is not named with \texttt{\_\_FREE} (line 6). (d) EXP file containing the quantitative data points shown in (b). The keyword \texttt{nan} is used to indicate missing data. (e) PROP file encoding the qualitative data points shown in (b). (f) CONF file used to configure PyBioNetFit. As described in the main text, the CONF file specifies the paths to the other files, the algorithm to be used, and the free parameters to be adjusted.}
\label{fig:demob}
\end{figure}

Figure \ref{fig:demob} shows an example set of PyBioNetFit input files for a simple problem. The problem is to parameterize a model for the chemical kinetics of three reactions (Fig. \ref{fig:demob}a) using (synthetic) quantitative and qualitative data (Fig. \ref{fig:demob}b). 

A model file (Fig. \ref{fig:demob}c; filename extension .xml for SBML models or .bngl for BNGL models)  defines the model to be fit. BNGL files may be prepared with a standard text editor, or with the text editor available within RuleBender \cite{Xu2011}, an integrated development environment for BioNetGen. SBML files are not human-readable, and should be prepared using SBML-compatible software such as COPASI \cite{Hoops2006} or Tellurium \cite{Medley2018,Choi2018}. Model files must conform to certain conventions for compatibility with PyBioNetFit. For BNGL files, each free parameter to be fit must be assigned a name ending in \texttt{\_\_FREE} (Fig. \ref{fig:demob}c lines 3-5) and each model output to be compared to measurements must be introduced as a BNGL observable or function. For SBML files, each free parameter must be an SBML parameter or the initial concentration of a species, and each model output to be compared to measurements must be an SBML species. In addition, each simulation command must have an associated string identifier, called a suffix. If the simulation is defined in a BNGL file, the suffix is specified using the \texttt{suffix} argument of the \texttt{simulate} or \texttt{param\_scan} action. If the simulation is defined in the CONF file (see description of the CONF file below), the suffix is specified as part of the \texttt{time\_course} or \texttt{parameter\_scan} declaration. The suffix must match the name of the corresponding experimental data file (e.g., ``d1'' in Fig. \ref{fig:demob}).  In the simplest case, a fitting job has one model file. However, PyBioNetFit supports jobs with multiple model files, such as, for example, the problem considered in Section \ref{sec:yeast}. This feature is useful when two or more models have parameters in common, such as two models that represent the same process in wild type and mutant cells.

Experimental data are supplied in EXP (Fig. \ref{fig:demob}d) and PROP (Fig. \ref{fig:demob}e) files. EXP files contain tabular data, such as time courses or dose-response curves. These files are specified in a space-delimited format in which the first column corresponds to the independent variable, and other columns correspond to dependent variables (the same format as is used in GDAT files output by BioNetGen). PROP files contain statements written using BPSL, which is described in the next section. 

A configuration file, or CONF file (Fig. \ref{fig:demob}f), provides the settings for running a fitting job. These settings include which model and data files to use (line 3), which parameters will be free to vary in fitting (lines 15-17), which fitting algorithm to use (line 9), which objective function to use (line 10), and settings specific to the selected fitting algorithm (lines 11-12) such as the mutation rate in the case of differential evolution. A CONF file may also include the keys \texttt{time\_course} and \texttt{param\_scan}, which are used to define simulation protocols. These keys are not needed for BNGL models, in which the same commands are available as BNGL actions (e.g., Fig. \ref{fig:demob}c lines 32-34). They are required for fitting SBML models because the model standard itself does not support definition of simulation protocols. A complete listing of the available configuration keys is provided in the PyBioNetFit user manual \cite{Mitra2019}.

After generating all of the required files, saved in the locations specified in the CONF file, a user can run PyBioNetFit from the command line with the command 
\begin{equation*}
    \texttt{pybnf -c } \textit{<filename>}\texttt{.conf}
\end{equation*} 
where \textit{<filename>} is the name of the CONF file. Figure \ref{fig:demoa}b illustrates the internal operations of PyBioNetFit. PyBioNetFit iteratively passes proposed parameter sets to the appropriate simulator, reads the simulation results, and calculates the value of the user-selected objective function. The objective function values are fed back into the optimization algorithm and affect which parameter sets are proposed in future iterations. Upon termination of the algorithm, PyBioNetFit outputs the best-fit parameter values, new model files that include those parameter settings, and (optionally) simulation results generated from those model files (Fig. \ref{fig:demoa}c).

\subsection{Property specification with BPSL}

To allow fitting to qualitative data, we implemented the approach described in ref. \cite{Mitra2018a} in PyBioNetFit. For this feature, we developed BPSL, a novel property specification language. BPSL is designed for writing system properties of cellular regulatory networks. In BPSL, system properties are expressed as inequalities involving the dependent variables of an experiment or model. We refer to such dependent variables in this section as simply ``variables.'' Typically, BPSL statements are written to use for parameterization of a particular model, in which case the names of the variables should match names of outputs of that model, similar to column headings of an EXP file. However, we note that like EXP files, BPSL statements primarily encode (experimental) data, and the same data could be considered in conjunction with any model for the system of interest (possibly only after changing the variable names to match the output names of the new model). Variables in BPSL are flexible: in addition to what is supported in EXP files---quantities corresponding to BNGL or SBML model outputs---it is possible to compare variables/readouts between different models/systems. One application of this feature would be case-control comparisons, such as comparing a mutant to wild type. 

Each inequality declared in BPSL is enforced at a particular value or range of values of the independent experimental variable (e.g., time). For example, an inequality might be enforced at one specific time, or at all times in a time course. As described below, BPSL syntax provides a means to define inequalities, where they are enforced, and how much they contribute to the objective function during optimization. 

A BPSL statement consists of three parts: an inequality, followed by an enforcement condition, followed by a weight. The inequality establishes a relationship (\texttt{<}, \texttt{>}, \texttt{<=}, or \texttt{>=}) between a variable and a constant or between two variables. The enforcement condition specifies where in a time course or dose-response curve the constraint is in effect. The enforcement condition is defined using the keywords \texttt{always}, \texttt{once}, \texttt{at}, and \texttt{between}, as summarized in Table \ref{table:const}. The weight (declared with the \texttt{weight} keyword) specifies the static penalty coefficient to be used during optimization when the inequality is not satisfied. Specifically, if the constraint $g(\textbf{x})<0$ is not satisfied with the trial parameter set $\textbf{x}$, the objective function adds a penalty equal to $C \cdot g(\textbf{x})$ where $C$ is the weight of the constraint. Note that, in this formulation, the penalty decreases as we move closer to satisfying the constraint. This feature of the objective function serves to guide an optimization algorithm toward constraint satisfaction.

We illustrate the use of BPSL with the following examples, assuming time-course outputs $X(t)$ and $Y(t)$. The BPSL statement \begin{equation*} 
\texttt{X>5 always weight 2} 
\end{equation*}
defines a constraint requiring $X(t)$ to be greater than 5 at all times. If the constraint is violated, a penalty of $2 \cdot (5-\textrm{min}(X(t)))$ is added to the objective function. The BPSL statement 
\begin{equation*}
\texttt{X<1 between time=8,Y=5 weight 3}    
\end{equation*}
defines a constraint requiring $X(t)$ to be less than 1 over a specified time range. The start point of this time range is specified directly: \texttt{time=8}. The end point is specified indirectly based on the value of $Y(t)$; it is at the first time point after $t=8$ where $Y(t)=5$. More precisely, to avoid numerical error, PyBioNetFit checks when $Y(t)$ \textit{crosses} 5, i.e.,  finds, after $t=8$, the first two consecutive output times $t_1$ and $t_2$ such that $Y(t_1)<5 \leq Y(t_2)$ or $Y(t_1)>5 \geq Y(t_2)$, and sets $t_2$ as the end point. If the constraint is violated at any point in the above time range, the penalty is $3 \cdot (\textrm{max}(X(t))-1)$, where $\textrm{max}(X(t))$ is evaluated over the above time range. 

\begin{table}[htb]
\centering
\begin{tabular}{|p{2.2cm}p{6cm}|}
Keyword & Meaning \\
\hline
\texttt{always} & At all times \\
\hline 
\texttt{once} & At one or more time points \\
\hline
\texttt{at} $\langle$\textit{condition}$\rangle$ & At the first time point where $\langle$\textit{condition}$\rangle$ is true \\
\hline
\texttt{between} $\langle$\textit{condition1}$\rangle$, $\langle$\textit{condition2}$\rangle$ & Over the range of time points starting with the first point where $\langle$\textit{condition1}$\rangle$ is true and ending with the first subsequent time point where $\langle$\textit{condition2}$\rangle$ is true \\
\hline
\end{tabular}
\caption{Keywords used to define enforcement conditions in BPSL. Definitions assume that the independent variable is time, but any arbitrary independent variable may be considered, as when considering a steady-state dose-response curve instead of a time course.}
\label{table:const}
\end{table}

\subsection{Metaheuristic fitting algorithms}

PyBioNetFit features four recommended parallelized metaheuristic fitting algorithms, which we will refer to as differential evolution (DE), asynchronous differential evolution (aDE), particle swarm optimization (PSO), and scatter search (SS). The details of each algorithm's implementation and configuration options are provided in the PyBioNetFit user manual \cite{Mitra2019}.

We tested the capabilities of these algorithms on a total of 31 example problems. Input files, descriptions, and results for each of these problems are provided in Supplementary File 1, a ZIP archive containing 31 numbered folders, one for each example problem. We will refer to the problems by these numbers, e.g., Supplementary File 1 (Problem 1). The problems feature models originally reported in refs. \cite{Kozer2013, Monine2010, Chylek2014a, Harmon2017, Erickson2019, Faeder2003a, Romano2014, Blinov2006, Kocieniewski2012, Mitra2018a, Dunster2014, Boehm2014,Brannmark2010,Zheng2012a,Fey2015,Webb2011,Mukhopadhyay2013,Lee2003,Suderman2013,Kuhn2016,Hlavacek2018,Shirin2018,Oguz2013}. Some of the models came from curated collections of models \cite{Thomas2016,Gupta2018,LeNovere2006,Hass2018}. In some cases, we fit models to published experimental data \cite{Kozer2013,Posner2007,Chylek2014a,Harmon2017,Kiselyov2009,Xue2000,Boehm2014,Brannmark2010,Zheng2012a,Fey2015,Manz2011,Yi2003,Yu2008,Leeuw1998,Spellman1998}. In other cases where no appropriate experimental dataset was available, we generated synthetic data by simulating the model with an assumed ground truth parameter set. The synthetic data included noise; depending on the problem, this was added as Gaussian (white) noise, uniformly distributed noise, or noise inherent to performance of a single stochastic simulation. All of the problems could be solved with an acceptable fit (defined as reaching a target objective function value, described in Supplementary File 1 for each individual problem) with at least one of the available algorithms using the default algorithmic parameters, and most could be solved with all four algorithms tested, albeit with different efficiencies. It is likely that efficiency could be improved with problem-specific choices for the algorithmic parameters, but we found that the default settings worked reasonably well for most of our example problems.  

To evaluate which algorithms are most effective in typical use cases, we performed timed benchmarking. We used the default algorithmic parameters for each algorithm. Because of the stochastic nature of the algorithms, many replicates of the same fitting job were necessary to make conclusions about the typical run time of each algorithm. We chose four benchmark problems with fitting run times on the order of hours. Such problems are not trivial, but it is still feasible to run many fitting replicates on a cluster. The selected benchmarks are described in Table \ref{table:bench}.

To examine the full scope of PyBioNetFit functionality, our selected benchmarks include one problem using each of the four simulators supported in PyBioNetFit, which we refer to as libRoadRunner, BioNetGen ODE, BioNetGen SSA, and NFsim. These simulators are described briefly as follows. (1) libRoadRunner is an SBML simulator. By default, libRoadRunner interfaces with CVODE \cite{Hindmarsh2005} to perform numerical integration. (2) BioNetGen ODE refers to the numerical integration capability of BioNetGen, accessed with the action \texttt{simulate(method=>"ode")}. Like libRoadRunner, this functionality interfaces with CVODE \cite{Hindmarsh2005}. (3) BioNetGen SSA refers to the stochastic simulation algorithm of BioNetGen, accessed with the action \texttt{simulate(method=>"ssa")}. (4) NFsim refers to the component of BioNetGen, accessed with the action \texttt{simulate(method=>"nf")}, that performs agent-based stochastic simulations without generation of a reaction network.

\begin{table}[!!!!!!!tb!!!!!!!!!!!!!!!!!]
\centering
\begin{tabular}{|p{2cm}|c|c|c|c|}
Benchmark & 1 & 2 & 3 & 4 \\
\hline
Description of model &\multicolumn{1}{|p{3.58cm}|}{Histone methylation at multiple sites} & \multicolumn{1}{|p{3.58cm}|}{EGFR-mediated activation of Grb2, Sos, and Shc} & \multicolumn{1}{|p{3.58cm}|}{Fc$\epsilon$RI signaling in response to receptor dimerization} & \multicolumn{1}{|p{3.58cm}|}{EGFR oligomerization and phosphorylation} \\
\hline
Format & SBML & BNGL & BNGL & BNGL \\
\hline
Simulator & libRoadRunner & BioNetGen ODE & BioNetGen SSA & NFsim \\
\hline
Rules & -- & 23 & 24 & 26\\
\hline
Reactions & 60 & 3749 & 58276 & Not calculated\\
\hline
Chemical species & 30 & 356 & 3744 & Not calculated\\
\hline
Free \mbox{parameters} & 46 & 37 & 20 & 9 \\
\hline
Output type & Time course & Time course & Time course & \begin{tabular}{@{}c@{}}Time course and \\ parameter scan\end{tabular} \\ 
\hline
Data type & Experimental & Synthetic & Synthetic & Experimental \\ 
\hline
Data points & 60 & 40 & 66 & 24 \\
\hline
Reference &\cite{Zheng2012a}& \cite{Blinov2006} & Original model: \cite{Faeder2003a} & \cite{Kozer2013} \\
&&&This version: \cite{Sneddon2011}& \\
\hline
\end{tabular}
\caption{Models used for timed PyBioNetFit benchmarking (Fig. \ref{fig:bench}). These models have all been used in earlier work to benchmark other software tools \cite{Hass2018,Gupta2018,Thomas2016}.}
\label{table:bench}
\end{table}

\begin{figure}[p!]
\centering
\includegraphics{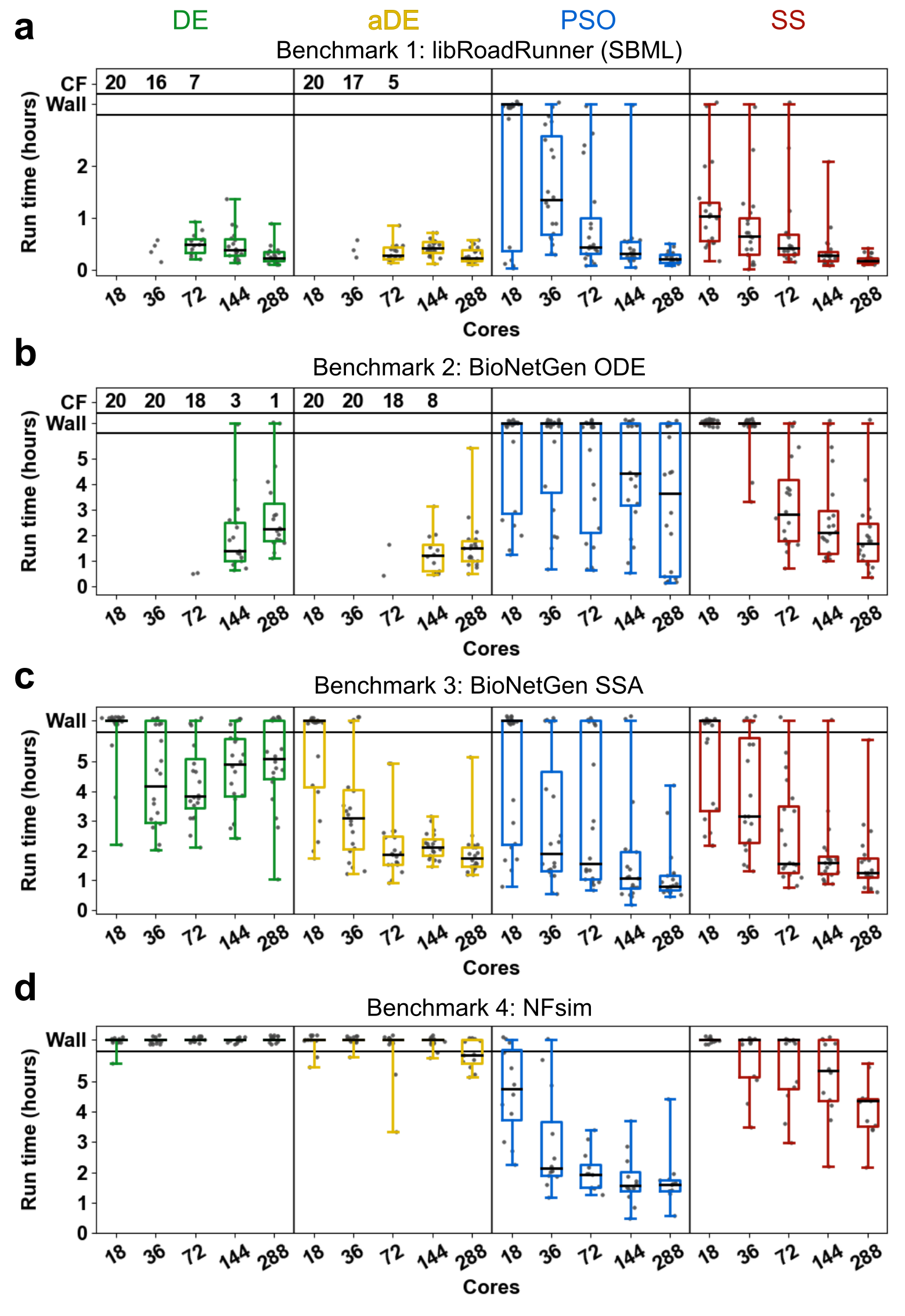}
\caption{Results from timed benchmarking of PyBioNetFit. Run times required to reach a target objective value are shown for the four benchmark problems described in Table \ref{table:bench}, for the DE, aDE, PSO, and SS algorithms implemented in PyBioNetFit. Box plots indicate the distribution of 20 replicates in a-c and 12 replicates in d. Gray points represent results from individual replicates. Replicates that ran for the maximum wall time (3 hours in a, 6 hours in b-d) without reaching the target value are plotted in the ``Wall'' band. When calculating percentiles for box plots, ``Wall'' replicates were taken to be larger than any successful replicate. ``CF'' (convergence failure) gives the number of replicates, out of 20 total, that failed because the population converged to a single point which was worse than the target value. Box plot statistics do not include convergence failures. Box plots are not shown for settings in which more than half the replicates were convergence failures.}
\label{fig:bench}
\end{figure}

For each of the benchmark problems, we chose a target objective function value as described in Methods, and measured the run times required for each algorithm to reach the target value. We evaluated the run times of the four algorithms, and also measured how the run times scaled with an increasing number of available cores on a cluster. The resulting distributions of run times are shown in Fig. \ref{fig:bench}. We found that in most cases, the algorithms show good capacity for taking advantage of parallelization, in that the median run time decreases as the number of available cores increases. The best algorithm varies by problem, and also varies by the number of cores available. Notably, with a large number of available cores (288), PSO (an asynchronous algorithm) is most effective for the benchmarks using stochastic simulators (BioNetGen SSA and NFsim) (Fig. \ref{fig:bench}c-d), but is outperformed by aDE and SS for the other benchmarks (Fig. \ref{fig:bench}a-b). 

\afterpage{\clearpage} 

Variability in algorithm performance is expected when considering a broad range of problems. In the end, the best algorithm and level of parallelization are problem-specific, and must be selected through trial and error. PyBioNetFit helps users in this regard by providing robust implementations of several algorithms, allowing for easy testing of different approaches. 

Two additional metaheuristic algorithms are implemented in PyBioNetFit but not rigorously benchmarked: simulated annealing (SA), and the parallelized island-based differential evolution (iDE) algorithm of ref. \cite{Penas2015a}. These algorithms were challenging to include in benchmarking due to the need to tune problem-specific parameters (temperature and step size in the case of SA, and tradeoffs between island size and number of islands versus the number of available cores in the case of iDE). Still, we include the implementations in PyBioNetFit with the hope that users will find them useful for specific problems. 

\subsection{Local optimization}

PyBioNetFit includes a parallelized implementation \cite{Lee2007} of the simplex algorithm \cite{Nelder1965}, a gradient-free local search algorithm. The simplex algorithm may be used on its own, or to refine the best fit obtained from any of the other algorithms, which is the use that we recommend. 

\subsection{Comparison to Data2Dynamics}

Although we did not rigorously benchmark PyBioNetFit against other parameterization tools, we tested the gradient-based method of Data2Dynamics \cite{Raue2015} on Benchmarks 1 and 2 (Table \ref{table:bench}), to obtain a rough view of how the performance of this tool compares to that of PyBioNetFit. Results are provided in Supplementary File 1 (Problems 11 and 19) and considered further in Discussion. 

\subsection{Uncertainty quantification}

\begin{figure}[tb!]
\centering
\includegraphics{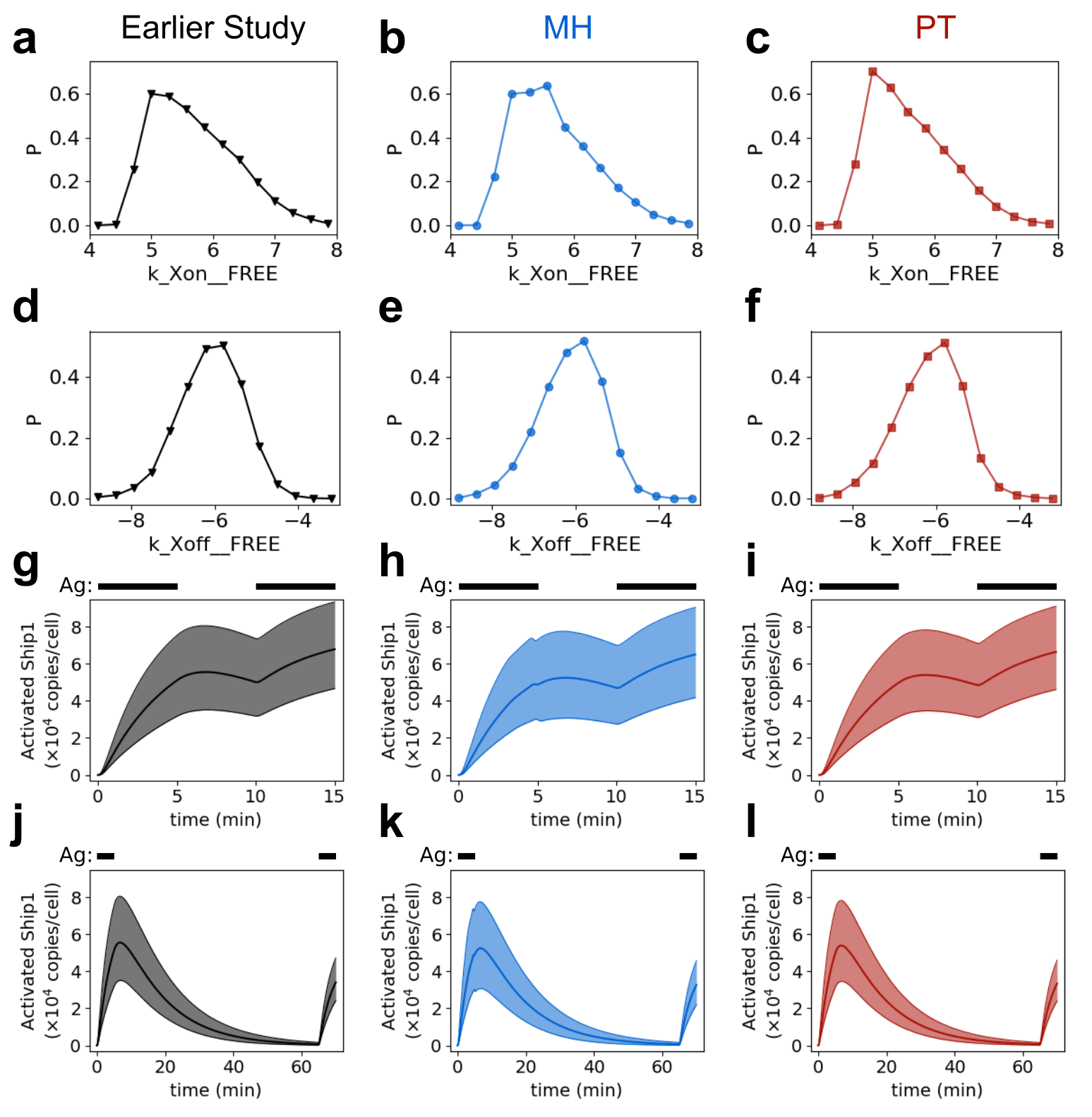}
\caption{Bayesian uncertainty quantification in PyBioNetFit. Results from PyBioNetFit's MH (b, e, h, k) and PT (c, f, i, l) algorithms are compared to the problem-specific code of ref. \cite{Harmon2017} (a, d, g, j). The data plotted in panels a, d, g, and j originally appeared in ref. \cite{Harmon2017}. (a-f) Marginal posterior probability distributions for selected parameters of the model of ref. \cite{Harmon2017}. Two examples out of the 16 model parameters are shown. (g-l) Prediction uncertainty quantification for time courses of activated Ship1, one of the model outputs.  Two antigen stimulation protocols are shown: one in (g-i) and the other in (j-l). Black bars above graphs indicate times when multivalent antigen was present. Solid curves indicate the median, and shaded areas indicate the 68\% credible interval.}
\label{fig:bayes}
\end{figure}

For Bayesian uncertainty quantification PyBioNetFit offers two MCMC methods: the conventional Metropolis-Hastings (MH) algorithm, and parallel tempering (PT). These methods are used by setting \texttt{fit\_type=mh} or \texttt{fit\_type=pt} in the CONF file, similar to how \texttt{de} is selected in Fig. \ref{fig:demob}f, line 9. To validate the accuracy of PyBioNetFit, we used MH and PT with the model of mast cell signaling described by Harmon et al. \cite{Harmon2017}. Harmon et al. \cite{Harmon2017} observed differences in mast cell degranulation as a function of the time delay between two pulses of antigen stimulation of IgE receptor activity. The model describes the activities of Syk and Ship1 during this two-stage stimulation protocol. The original study included Bayesian uncertainty quantification of model parameters and predictions using problem-specific code that implemented MH. We ran MH and PT in PyBioNetFit using input files provided in Supplementary File 1 (Problem 7). We found that both PyBioNetFit algorithms achieved good agreement with the published results for parameter uncertainty (Fig. \ref{fig:bayes}a-f) and prediction uncertainty (Fig. \ref{fig:bayes}g-l). For this problem, the MH and PT algorithms converged to the correct distribution at roughly the same rate. 

Note that to calculate the posterior probability distribution, the objective function is assumed to represent a log likelihood. This assumption is valid for the chi squared objective function, which was used in this example. We do not recommend using Bayesian MCMC algorithms with PyBioNetFit's other available objective functions, or with contributions from qualitative data added to the objective function.

When using Bayesian MCMC methods, it is important to choose algorithmic parameters such that the posterior distribution is sampled accurately. In particular, some number of unsampled ``burn-in'' iterations should be used to allow the Markov chains to reach a starting point in a region of high likelihood. In addition, an adequately large number of iterations must be sampled for the Markov chains to fully explore the posterior distribution. Exploring the distribution may be especially challenging when the distribution is multimodal, and it is a rare event for a Markov chain to move between modes. In these situations, PT is expected to outperform MH by providing a faster means of escape.

Run times of Bayesian MCMC algorithms are expected to be dominated by the run times of the large number of simulations required to adequately sample probability distributions. We therefore do not expect a noticeable difference in performance between different  implementations of the same MCMC algorithm run with the same settings, aside from differences in simulator efficiency. PyBioNetFit is a convenient tool for running MCMC because it supports both BNGL and SBML models without the need for custom code. In addition, MCMC in PyBioNetFit takes advantage of parallelization. In MH, individual Markov chains are not parallelizable, but PyBioNetFit can run multiple independent Markov chains in parallel and pool the results to create a larger sample of a probability distribution. In PT, the algorithm requires the simultaneous propagation of several Markov Chains, and these chains are run in parallel.

PyBioNetFit offers bootstrapping \cite{Efron1994,Press2007} as another uncertainty quantification method. Bootstrapping is configured by adding the line \texttt{bootstrap=}$\langle$\textit{number}$\rangle$ to the CONF file, where $\langle$\textit{number}$\rangle$ is the number of bootstrap replicates to perform. Bootstrapping relies on an assumption that the experimental data points are drawn from some (unknown) probability distribution, and drawing a sample from the data available is a good approximation of drawing a sample from the distribution.

Bootstrapping provides a different type of metric from Bayesian methods: it provides a confidence interval specifically for the location of the \textit{best-fit} parameter set. Importantly, this confidence interval depends on both the experimental data and the performance of the fitting algorithm used, unless the fitting algorithm has perfect performance with respect to finding the global minimum. When we obtain a ``90\% confidence interval'' from bootstrapping, it means that if the experiment was repeated, and then the fitting was repeated using the new data, then the best-fit parameter is expected to fall within the interval with 90\% confidence.

To illustrate how bootstrapping can be used to measure uncertainty arising from both data and fitting algorithms, we consider a fitting problem consisting of an egg-shaped curve (Supplementary File 1, Problem 27), originally presented in ref. \cite{Hlavacek2018}. This toy problem is simple enough for PyBioNetFit's SS algorithm to find the global minimum, but the BioNetFit 1 algorithm is less effective. We performed bootstrapping on this problem with PyBioNetFit, and found that the best fit for each parameter was identified to precision of order $10^{-4}$ with 90\% confidence (Supplementary File 1, Problem 27). This high level of precision is unsurprising, given that the input data consist of densely sampled points on the target curve with minimal noise. In contrast, 90\% confidence intervals reported using BioNetFit 1 span large ranges (of order 1 in some cases) \cite{Hlavacek2018}. We conclude that the uncertainty reported with BioNetFit 1 arises mainly from limitations of the fitting algorithm, rather than limitations in the amount or quality of data for fitting. Again, results from bootstrapping would be independent of the optimizer if the optimizer is always able to find the global minimum, but this is not a realistic expectation for many problems.

In summary, the uncertainty quantification methods in PyBioNetFit provide different and complementary functionalities. The Bayesian MH and PT algorithms estimate a multidimensional probability distribution showing the most probable parameters (treated as random variables) based on the data. Different Bayesian algorithms with the same input data are expected to produce the same results, as long as algorithmic settings allow for sufficient sampling of the posterior probability distribution.  Bootstrapping evaluates the uncertainty given a (metaheuristic) fitting algorithm in combination with a particular data set. The resulting bootstrap confidence interval represents the confidence in the best-fit parameter values if both the experiment and the fitting were to be repeated.

\subsection{Application: fitting a model of yeast cell cycle control using both qualitative and quantitative data}
\label{sec:yeast}

To demonstrate the capabilities of PyBioNetFit to parameterize models using both qualitative and quantitative data, we used PyBioNetFit to solve a challenging problem that we previously solved with problem-specific code \cite{Mitra2018a}. We parameterized the model of yeast cell cycle control described in refs. \cite{Oguz2013,Laomettachit2011}, incorporating the qualitative data tabulated in ref. \cite{Laomettachit2016} and the quantitative data in ref. \cite{Spellman1998}. The full statement of the problem and the objective function are the same as in ref. \cite{Mitra2018a}. 

PyBioNetFit contains all of the features needed to repeat the fitting job of ref. \cite{Mitra2018a}. The input files to run the fitting job are provided as Supplementary File 1 (Problem 30). Like in the original study, we performed optimization using scatter search, as described in Methods. 

\begin{figure}[tbp!]
\centering
\includegraphics{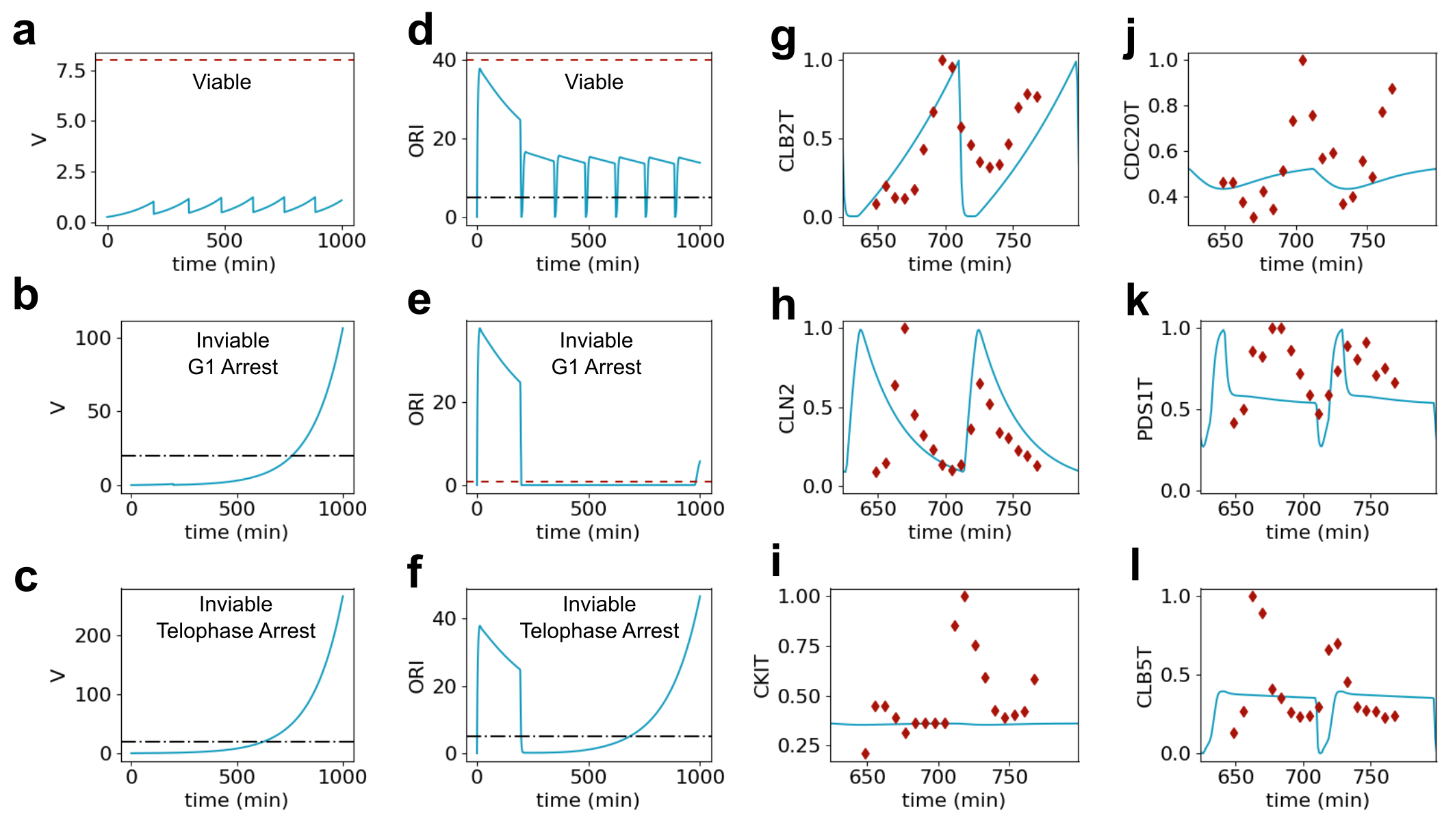}
\caption{Example outputs of the model for yeast cell cycle control parameterized with PyBioNetFit. (a-f) Selected output showing agreement with qualitative data. Two output variables are shown: $V$ (a-c), representing cell volume and $ORI$ (d-f), a flag that indicates origin activation is completed when its value reaches 1. Results for three selected yeast strains are shown: wild type (a,d), which is viable, a mutant (\textit{cln3$\Delta$ bck2$\Delta$}) (b,e), which has a G1 arrest phenotype, and another mutant (\textit{cdc14-ts}) (c,f), which has a telophase arrest phenotype. Horizontal lines indicate qualitative constraints: time courses should exceed black dash-dot lines, and should not exceed red dashed lines. (g-l) Selected output showing agreement with quantitative data of ref. \cite{Spellman1998} (red diamonds). These plots were shown in ref. \cite{Mitra2018a} with the best-fit results obtained in that study. }
\label{fig:yeast}
\end{figure}

We ran the fitting job in PyBioNetFit, and present a subset of the results in Figure \ref{fig:yeast}, and the parameterized model in Supplementary File 1 (Problem 30). We achieved a minimum objective function value of 80, compared to 70 in the original study, a difference which is not surprising given the stochastic nature of the SS algorithm (or any metaheuristic). Our fit is not identical to the previously published fit, which is expected because some model parameters were shown not to be identifiable \cite{Mitra2018a}. However, like the published fit, the fit generated by PyBioNetFit shows reasonable consistency with the qualitative data (Fig. \ref{fig:yeast}a-f) and the quantitative data (Fig. \ref{fig:yeast}g-l). 

\subsection{Applications beyond fitting: model checking}

\begin{figure}[tbp!]
\centering
\includegraphics{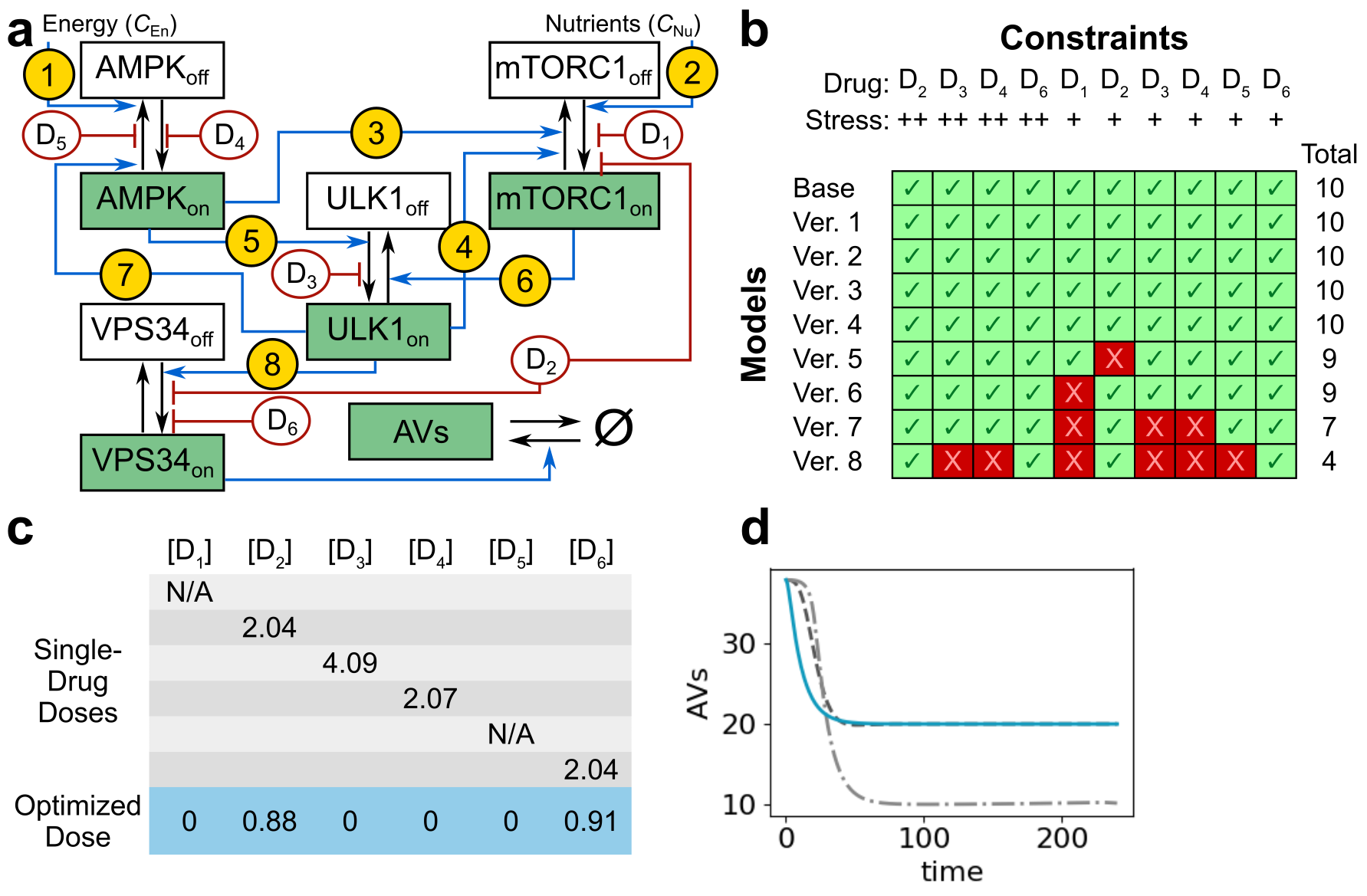}
\caption{Applications of PyBioNetFit in analysis of a parameterized model. (a) Schematic of the model to be analyzed, adapted from ref. \cite{Shirin2018}. Six drugs labeled $\textrm{D}_1$ through $\textrm{D}_6$ are capable of modulating various processes in the network shown. Interactions among kinases considered in the model are numbered 1--8. (b) Model checking performed by PyBioNetFit of hypothetical alternatives to the model shown in (a). Each alternative model version (numbered 1--8) was obtained by removing one interaction from (a), corresponding to the version number. Each model version was checked against ten qualitative behaviors characterized in ref. \cite{Shirin2018}: the change (increase or decrease) in AV count in response to a particular drug at a particular stress level, high (++) or medium (+). (c) Optimizing drug dosing to achieve a desired system behavior. This table shows the minimal constant drug concentration to reduce AV count to 20 per cell or below, in a cell under high stress. Gray rows show optimized doses for each drug individually. The blue row shows the optimized dose from simultaneously tuning all six drug concentrations. Although all six drug doses were allowed to vary, the optimal solution had only two drugs with nonzero dose. (d) Time course of AV counts under the treatments shown in (c). The gray broken line shows the response to treatment with $\textrm{D}_3$ only, the gray dash-dot line the response to treatment with $\textrm{D}_4$ only, and the blue solid line shows the response to the optimized six-drug dose. Responses to  $\textrm{D}_2$ or $\textrm{D}_6$ only are indistinguishable from the response to the optimized dose (but require more total drug than with the optimized drug combination). }
\label{fig:extra}
\end{figure}

Although PyBioNetFit was designed for model parameterization, the property specification language of PyBioNetFit has additional applications in the analysis of parameterized models, namely model checking and design. To demonstrate these applications, we consider the model of Shirin et al. \cite{Shirin2018} (Fig. \ref{fig:extra}a). The model describes the interactions between four kinases involved in regulation of autophagy, a cellular recycling process. The model also describes the effects of six specific drugs in modulating these interactions and the level of autophagy. In the original study, this model was used to investigate the capabilities of the six drugs (labeled $\textrm{D}_1$ through $\textrm{D}_6$) to control the number of autophagic vesicles (AVs) in a cell. For our analysis, we assume that the published parameterization of the model, which was shown to be consistent with certain experimental data in the original study, is acceptable.

We use this model to illustrate PyBioNetFit's model checking capabilities. In general, model checking consists of evaluating whether a particular model satisfies a set of specified properties \cite{Clarke1999}. In biological modeling, we might have multiple possible models designed to describe the same biological system, and we want to know which model is most consistent with the known properties of that system.

For the autophagy model of Shirin et al. \cite{Shirin2018}, we consider eight hypothetical alternative models, each obtained by removing one of the labeled interactions from the network illustrated in Fig. \ref{fig:extra}a. These changes are arbitrary for demonstration of the model checking workflow, but represent a scenario that could arise in practice: often, many models of the same biological process are developed by different research groups for different purposes, and a particular interaction might be present in one model but absent in another. In such a scenario, it is reasonable to ask whether the interaction is important to the model's ability to reproduce certain system properties. As system properties to be checked in our demonstration, we use the characterization of the system's response to drug treatment by Shirin et al. \cite{Shirin2018}, which we take to be the established truth. Specifically, for the six drug treatments allowed in the model and two levels of cellular stress (controlled by the energy and nutrient parameters of the model, $C_{\textrm{En}}$ and $C_{\textrm{Nu}}$), Shirin et al. \cite{Shirin2018} characterized the change in number of AVs relative to control. Ten of these 12 model settings resulted in an increase or decrease in AV count. For our model checking exercise, we determined whether each of our hypothetical alternative models is able to reproduce each of these ten qualitative behaviors. 

PyBioNetFit includes a model checking utility for checking properties in BPSL. Each property of interest for this model can be written in BPSL as an inequality between the AV count for the untreated case and the AV count for the drug treated case. In PyBioNetFit, this type of case-control comparison is configured by performing simulations corresponding to multiple versions of the model---here, one version for each stress/drug combination considered, plus one version at each stress level with no drug. As described in Section \ref{sec:config}, PyBioNetFit requires each of these simulations to have a unique suffix (a string defined in the BNGL or CONF file). These suffixes can be used in the BPSL file to refer to the outputs of specific simulations. For example, suppose that the simulation of wild type has the suffix \texttt{data}, the simulation with drug $\textrm{D}_2$ has suffix \texttt{data\_D2}, and after the system has equilibrated (defined as the time window 120--240 minutes in the time course), the AV count should be lower in the presence of $\textrm{D}_2$. Then the constraint would be written as:
\begin{equation*}
\texttt{data.AV > data\_D2.AV between 120,240}    
\end{equation*}
The full implementation of the model checking problem in PyBioNetFit is provided as Supplementary File 1 (Problem 31). 

The results of model checking are shown in Fig. \ref{fig:extra}b. Four of the variant models (versions 1--4) remain consistent with all ten system properties, whereas the other four (versions 5--8) no longer satisfy one or more of the properties. In the context of this model, these results suggest that interactions 1--4 in Fig. \ref{fig:extra}a are not essential to the qualitative properties that we considered. More generally, this example demonstrates the ability of PyBioNetFit's model checking utility to help distinguish between a set of models. 

\subsection{Applications beyond fitting: design}

PyBioNetFit can also be used to solve design problems in which we seek a perturbation of a system that produces a specified system response. In such a problem, the desired behavior is defined in BPSL. We again examine the autophagy model of Shirin et al. \cite{Shirin2018} and its published parameterization, and consider a similar problem to the original study. We want to choose the concentrations of drugs $\textrm{D}_1$ through $\textrm{D}_6$ so as to drive the AV count below a desired threshold, while minimizing the total quantity of drug used. We arbitrarily choose a threshold of 20 AVs, and set $C_{\textrm{En}}=C_{\textrm{Nu}}=0.1$ (on a scale of 0 to 1), corresponding to a high level of cellular stress. In the original study, arbitrary time courses of drug dosing were permitted, and simultaneous dosing of up to two drugs at a time (out of the six drugs in the model) was considered. Here, we solve a different problem in which we limit ourselves to constant drug concentrations, but allow for simultaneous dosing of up to six drugs. 

We configure this problem as a fitting problem in PyBioNetFit, in which the free parameters to be fit represent the unknown concentrations of each of the six drugs. The desired system property of reducing AV count below 20 is implemented as an inequality constraint, and the goal to minimize drug concentration is implemented as a quantitative data point (i.e., minimizing the difference between the actual total drug dose and 0). The full configuration of this problem is provided as Supplementary File 1 (Problem 31).

The optimization results are shown in Fig \ref{fig:extra}c (bottom row).  For comparison, we also performed optimizations in which only one of the drug concentrations was allowed to vary (Fig. \ref{fig:extra}c). The results are consistent with those reported in ref. \cite{Shirin2018}. Note that the optimized combined dose allowing all six drugs uses less total drug than any of the single-drug doses. The optimized dosing schemes for both single-drug and combination treatments achieve the desired property of driving AV count below 20 (Fig. \ref{fig:extra}d).

This example illustrates an additional, important class of problem that can be addressed with PyBioNetFit: the design of perturbations to a biological system to achieve specified behavior. In particular, optimization of therapeutic treatments has been a long-standing goal in the systems biology field \cite{Fitzgerald2006}. The automated design of perturbations, with formal definition of target behavior, is systematic and less likely to miss effective perturbations than an \textit{ad hoc} approach to model analysis. 

\section{Discussion}

\subsection{Comparison to related tools}

Some features of PyBioNetFit are unique and novel, while other features have some overlap with other available optimization tools. Here we analyze and discuss the strengths and weaknesses of PyBioNetFit in comparison to other published tools. 

For parameterization of models defined in BNGL, to our knowledge the only out-of-the-box general-purpose tools are PyBioNetFit and its predecessor BioNetFit 1 \cite{Thomas2016}. PyBioNetFit makes major improvements over BioNetFit 1 in terms of new functionality, as well as improved implementation of BioNetFit 1 functionality. 

In our experience, PyBioNetFit far outperforms BioNetFit 1. As one example comparison, we ran Benchmark 2 in BioNetFit 1 with population size 144 (Supplementary File 1, Problem 11). We considered using the cluster computing capabilities of BioNetFit 1, but found that the fitting ran faster on a single node (due to overhead in communicating with the cluster manager). BioNetFit 1 was unable to reach the target objective value within 10 hours in any of 5 fitting replicates. For comparison, the fastest PyBioNetFit algorithm at a parallel count of 144 on a cluster had a median run time of 1.9 hours (Fig. \ref{fig:bench}b).

A larger set of software is available for parameterization of ODE models defined in SBML. We considered Data2Dynamics \cite{Raue2015}. Data2Dynamics requires ownership of commercial software (MATLAB), but is otherwise open source. Data2Dynamics includes optimization functionality for ODE models, but does not directly support stochastic models or constrained optimization. Data2Dynamics takes advantage of parallel computing if a single evaluation requires simulation of multiple models, but the optimization algorithm itself is not parallelized. Data2Dynamics differs from PyBioNetFit in that its primary optimization algorithm is gradient-based (using the MATLAB function lsqnonlin, which implements a trust region reflective algorithm \cite{MathWorks2018}). Although Data2Dynamics also offers some metaheuristic methods, the lack of parallelism limits their efficiency for difficult problems, and the authors recommend lsqnonlin as the strongest algorithm in the suite. 

The default Data2Dynamics algorithm incorporates the method of sensitivity equations of ref. \cite{Leis1988}. In this method, the ODE system is augmented with additional variables and associated equations which track gradient information required for a gradient-based optimization algorithm
\cite{Raue2013}.


In PyBioNetFit, we chose not to include gradient-based methods at this time because existing tools, including Data2Dynamics, already provide acceptable solutions, and instead focused on metaheuristic optimization algorithms. Because of the different classes of algorithms used, PyBioNetFit and Data2Dynamics are best suited for different types of problems. Raue et al. \cite{Raue2013} showed that for two example fitting problems with small numbers of chemical species (6 and 26 respectively, comparable to our Benchmark 1), gradient-based optimization with lsqnonlin outperformed metaheuristic optimization algorithms. We compared the performance of Data2Dynamics on two of our four benchmark problems, Benchmarks 1 and 2 (Table \ref{table:bench}). We focused on Benchmarks 1 and 2 because Data2Dynamics does not support stochastic models (such as Benchmarks 3 and 4). 

We found that Data2Dynamics solves Benchmark 1 (Table \ref{table:bench}) in under 5 minutes (Supplementary File 1, Problem 19), significantly outperforming PyBioNetFit for this problem. This benchmark problem is well-suited for solution via gradient-based optimization because the model consists of a relatively small ODE system. Additionally, the parameter space appears not to be rugged, so gradient-based methods are not likely to become trapped in a local minimum. This result is consistent with the findings of Raue et al. \cite{Raue2013} for problems of similar difficulty.

We also considered Data2Dynamics for Benchmark 2 (Table \ref{table:bench}) (Supplementary File 1, Problem 11). Data2Dynamics does not directly support BNGL files, but we used a script to convert the reaction network generated by BioNetGen into a Data2Dynamics file with 356 chemical species and the same number of differential equations. We found that the gradient-based method of Data2Dynamics did not perform well on this model: the total run time was over 7 hours (run on a single core, because this Data2Dynamics method does not benefit from multiple cores), and the objective function value of the final fit was very poor (orders of magnitude above the target objective value). For comparison, PyBioNetFit reached the target objective value with a median runtime of 1.9 hours using its fastest algorithm (for this problem) on 144 cores. This fitting problem illustrates the limitations of gradient-based methods. First, for rule-based models, the ODE system implied by the rules can easily become high dimensional, especially after augmenting the system to include sensitivity equations. Such a system is computationally expensive for numerical integration, which empirically scales as the cube of the number of equations. Second, gradient-based optimization is susceptible to becoming trapped in local minima with poor objective function values. This issue can be partially remedied by performing multiple optimization runs at different start points, but can become prohibitive if the parameter space has too many local minima.  

The performance of Data2Dynamics on Benchmarks 1 and 2 suggests that gradient-based optimization and metaheuristic optimization perform well on different types of problems. PyBioNetFit can therefore be thought of as a complementary tool to the existing tools that support gradient-based optimization. 

COPASI \cite{Hoops2006} also supports parameterization of SBML models using either gradient-based or metaheuristic optimization algorithms. COPASI has a separately installed extension that enables cluster computing \cite{Kent2012}, including support for multistart optimization, but does not support parallelism within a single optimization run. Although we did not rigorously benchmark COPASI, we expect it to perform well in parameterizing small models (e.g., tens of chemical species), with gradient-based methods being the most efficient, as in ref. \cite{Raue2013}. However, like Data2Dynamics, its metaheuristic methods are limited by a lack of parallelism, and its parameterization functionality has not been demonstrated on large ODE systems such as those implied by rule-based models. 


PyBioNetFit is unique in its support for property specification with BPSL. Unlike previous work on biological property specification \cite{Clarke2008,Heath2008,Kwiatkowska2008,David2012,Liu2017,Hussain2015,Khalid2018}, which relied on problem-specific codes, BPSL can be used with the general-purpose functionality of PyBioNetFit. BPSL is also designed to be more human readable than conventional linear temporal logic (LTL), for instance. We expect a BPSL statment (but not an LTL expression) to be understandable to anyone with a background in biological modeling. For example, consider the following BPSL statement:
\begin{equation*} 
\texttt{A<1 between B=2,B=3}
\end{equation*}
This statement is equivalent to the following LTL expression:
\begin{equation*} 
\textbf{F}(B=2)\implies((\neg(B=2)) \textbf{U} (B=2 \wedge (A<1 \textbf{W} B=3)))
\end{equation*}
where $\textbf{F}$ is the ``future'' operator, $\textbf{U}$ is the ``until'' operator, and $\textbf{W}$ is the ``weak until'' operator. A drawback of BPSL relative to LTL is that the available enforcement keywords (Table \ref{table:const}) enable only a subset of what is possible with LTL. However, the current BPSL grammar is sufficient to support all constraints formulated in ref. \cite{Mitra2018a} to fit the yeast cell cycle model of refs. \cite{Oguz2013,Laomettachit2011}. Furthermore, PyBioNetFit was written with extensibility in mind, such that it is possible to add to the BPSL grammar as needs arise in other modeling problems.

\subsection{Comparison to problem-specific coding}

PyBioNetFit joins Data2Dynamics and COPASI in the class of domain-specific software supporting biological model parameterization, and has strengths that are complementary to these existing tools. These domain-specific tools contrast with the approach of using problem-specific code written in a high-level programming language such as Python, R, or MATLAB. We acknowledge that problem-specific code is a good choice in some use cases, such as when the model of interest is already implemented in one of these languages, or contains some unusual feature that is not supported in SBML or BNGL. Many packages are available that can help streamline model parameterization in high-level programming languages. For coding a model, standard differential equation packages could be used, or PySB \cite{Lopez2013} could be used to define a model in a form similar to BNGL or SBML. Other packages implement metaheuristic optimization algorithms \cite{Egea2014,Garrett2012,Fortin2012} and Bayesian uncertainty quantification algorithms \cite{Eydgahi2013,Gupta2018a,Shockley2018}. Packages such as dask.distributed \cite{Rocklin2015} are available to help with parallelization on clusters. 

Even with this array of tools, some amount of custom code is necessary to use these disparate packages to solve a given problem of interest.  We argue that in cases where writing BNGL or SBML models is feasible, the domain-specific functionality of PyBioNetFit is preferable to problem-specific code. PyBioNetFit combines all the functionality required for model parameterizaton into a single package. It removes the need for debugging at the level of the programming language, which reduces the propensity for errors in the modeling work. PyBioNetFit allows a modeler to instead focus on designing models and choosing appropriate algorithms for parameterization and analysis.  A second advantage of using PyBioNetFit is in reproducibility of results \cite{Medley2016,Waltemath2016}. Although it is possible to create well-documented, reproducible problem-specific code, such as in an IPython or R notebook, such good practices are not always followed. In contrast, a PyBioNetFit job can be documented by simply providing the set of input files used. The input files make it immediately clear to a reader what model, data, and algorithm configuration were used in the analysis.  

\subsection{Conclusion}

PyBioNetFit offers a versatile set of tools, which we expect to be useful in parameterization of new biological models. PyBioNetFit is most useful when the model of interest is expressed in SBML or BNGL, and when metaheuristic algorithms are preferable to gradient-based algorithms (because gradient information is expensive or unavailable, or the parameter landscape has many local minima). PyBioNetFit is also the first available implementation of our recent approach \cite{Mitra2018a} for leveraging both quantitative and qualitative data in fitting. This approach is enabled by BPSL, which can also be used for model checking and design. 

Our hope is that PyBioNetFit lowers the technical barrier to parameter fitting, by enabling fitting without problem-specific coding. PyBioNetFit will also promote reproducible modeling by encouraging the use of existing model standards (BNGL and SBML), and providing standardized algorithms with usage that can be specified unambiguously through PyBioNetFit's input files. 

We have shown that parameter identification can be challenging, and the best choice of fitting algorithm is not always obvious. By providing robust implementations of several algorithms, we encourage experimentation with different algorithms and settings to find the best choice for a problem of interest. 

\section{Methods}

\subsection{Software availability}
The most recent version of PyBioNetFit is v1.0.0, available online at \url{https://github.com/lanl/PyBNF}. The repository includes a user manual, \texttt{Documentation\_PyBioNetFit.pdf}, and the same user manual is available online as a standalone web site \cite{Mitra2019}. General information about PyBioNetFit is available at \url{http://bionetfit.nau.edu/}.

PyBioNetFit can be installed on any current Linux or macOS operating system. Installation of Python 3 is required if it is not already included with the operating system. PyBioNetFit can be installed from source by downloading the code at the above GitHub link, or can be installed directly using the pip package manager with the command 
\begin{equation*}
    \texttt{python3 -m pip install pybnf}
\end{equation*}

The example fitting problems provided in Supplementary File 1 are also available online at \url{https://github.com/RuleWorld/RuleHub/tree/master/Contributed/Mitra2019}.

\subsection{Implementation details}

PyBioNetFit is written in Python 3.6. The PyBioNetFit package includes novel code, and functionality that is provided through installation of dependencies. The following features are implemented by novel code: all supported optimization algorithms, parsing of CONF, EXP, and PROP files, processing of simulation results including evaluating a user-selected objective function. Full documentation of the PyBioNetFit code and its features is provided in the user manual \cite{Mitra2019}.

Implementations of optimization algorithms are based on published descriptions of the algorithms. Myriad variants of the algorithms have been described in the literature; here we cite the specific descriptions that we referred to when implementing the algorithms in PyBioNetFit. The iDE algorithm is described in ref. \cite{Penas2015a}, and our DE and aDE algorithms are based on the simpler algorithm described in the same reference. Our implementation of PSO is based on ref. \cite{Moraes2015}. Our implementation of SS follows the outline presented in the introduction of ref. \cite{Penas2017}, and uses the recombination method of ref. \cite{Egea2009}. Our implementation of MH MCMC  is described in ref. \cite{Kozer2013}. Our implementation of PT is described in ref. \cite{Gupta2018a}. Our implementation of SA is analogous to MH, but with a temperature parameter that decreases over the course of the run. Our parallelized implementation of the simplex algorithm is described in ref. \cite{Lee2007}. All algorithms are described in full detail in the PyBioNetFit user manual \cite{Mitra2019}.

PyBioNetFit interfaces with the dask.distributed \cite{Rocklin2015,Dask2018} package to provide parallelization on multi-core workstations or computer clusters. PyBioNetFit submits simulation jobs to dask.distributed, and dask.distributed assigns those jobs to the available workers as efficiently as possible. Dask.distributed is automatically installed during installation of PyBioNetFit. The user does not typically need to interact directly with dask.distributed, except when using certain unusual cluster environments (e.g., clusters in which SSH access between nodes is only possible with host-based authentication). 

PyBioNetFit uses third-party software to run simulations of models. libRoadRunner \cite{Somogyi2015} is a Python package used to run SBML models. libRoadRunner is automatically installed during installation of PyBioNetFit and does not require manual configuration. BioNetGen \cite{Harris2016} is used to run BNGL models; it provides the simulation methods described in Results as BioNetGen ODE, BioNetGen SSA, and NFsim. BioNetGen must be installed manually as a dependency of PyBioNetFit. The path to BioNetGen must be provided to PyBioNetFit using the \texttt{bng\_command} key in the CONF file or by setting the environment variable \texttt{BNGPATH}. Complete installation instructions are provided in the user manual \cite{Mitra2019}. 

\subsection{Running benchmark problems}
Example fitting problems were run in PyBioNetFit using the model, data and configuration files provided in Supplementary File 1, using a variety of computing resources as described in the README files in Supplementary File 1. 

Timed benchmark problems (Fig. \ref{fig:bench}) were run on a homogeneous computer cluster, consisting of Intel E5-2695\_v4 nodes. Each node had 36 cores, 125 GB RAM, and a base clock rate of 2.10 GHz. The nodes supported multithreading (2 threads per core), but PyBioNetFit was configured such that only one worker process per core was created. To run benchmarks with a specified number of cores, PyBioNetFit v0.2.2 was run with the appropriate number of nodes allocated (benchmarks using 36, 72, 144, and 288 cores were run on allocations of 1, 2, 4, and 8 nodes respectively). For 18-core benchmarks, two PyBioNetFit processes with 18 worker processes each were simultaneously run on the same 36-core node. 

Input files required to run the four benchmarks are provided in Supplementary File 1 (Problems 2, 9, 11 and 19). The \texttt{parallel\_count} and \texttt{population\_size} settings in the provided CONF files were edited to equal the number of available cores for the run. For SS, because the number of parallel simulations for population size $n$ is $n(n-1)$, \texttt{population\_size} was instead set to 4, 6, 9, 11, and 17 for core counts of 18, 36, 72, 144, and 288 respectively. 

For Bayesian uncertainty quantification (Fig. \ref{fig:bayes}), we used PyBioNetFit to sample 270,000 parameter sets by MH and 54,000 parameter sets by PT (configuration provided in Supplementary File 1 (Problem 7)). Note that the PT run used the same total number of simulations, but obtained fewer samples because only the replicas at the lowest temperature are sampled. To compare with the results of Harmon et al. \cite{Harmon2017}, we used the raw list of 270,000 samples generated in the original study. Raw data from all three runs are available in the BioStudies database (\url{http://www.ebi.ac.uk/biostudies}) under accession number S-BSST240. We generated histograms (Fig. \ref{fig:bayes}a-f) directly from the lists of sampled parameters. To generate prediction uncertainty plots (Fig. \ref{fig:bayes}g-l), we reran simulations using each sampled parameter set, and plotted the median, and the 16th and 84th percentile values, at each time point. We note that the 16th and 84th percentile curves were plotted erroneously in the original study, and are accurate in the present work.

Fitting of the yeast cell cycle control model (Fig. \ref{fig:yeast}) consisted of 40 fitting replicates, run on a heterogeneous set of nodes on multiple computer clusters. Each replicate used 2 nodes each with 28 to 44 cores, and completed in under 48 hours. Model, data, and configuration files required to run this problem are provided in Supplementary File 1 (Problem 30). We configured the fitting job using SS, and kept the same algorithmic settings as the original study with a few exceptions. Whereas the original study performed a single fitting job for 70,000 iterations, we ran an ensemble of 40 replicates of the job for 5000 iterations each. The modified protocol is a reasonable alternative because, in the original study, the fitting did not make steady progress throughout the run, but rather spent most of the time (all time after iteration 500) searching through fairly good parameter space in search of an extremely good parameter set. The modified protocol is better able to take advantage of parallelism: with sufficient parallel resources, all replicates can be run in parallel, and the entire fitting job can be completed in about 30 hours, compared to about 10 days in the original study. We also added a short, 500-iteration refinement of the best fit using the simplex local search algorithm. 

To analyze the autophagy model (Fig. \ref{fig:extra}) we implemented the model of ref. \cite{Shirin2018} in SBML format using COPASI \cite{Hoops2006}. We ran PyBioNetFit on the input files provided in Supplementary File 1 (Problem 31), which includes the SBML model file.

\section*{Acknowledgements}

This work was supported by grant R01GM111510 from the National Institute of General Medical Sciences (NIGMS) of the National Institutes of Health (NIH). WSH acknowledges support from the Joint Design of Advanced Computing Solutions for Cancer (JDACS4C) program established by the U.S. Department of Energy (DOE) and the National Cancer Institute (NCI) of NIH. RS and AI acknowledge support from the Center for Nonlinear Studies at Los Alamos National Laboratory (LANL), which is operated for the National Nuclear Security Administration (NNSA) of the DOE under contract 89233218CNA000001. HMS acknowledges the support of grants R01GM123032-01 from NIGMS/NIH and P41EB023912 from the National Institute of Biomedical Imaging and Bioengineering (NIBIB) of NIH. We thank J. Kyle Medley and Kiri Choi for assistance with libRoadRunner development. Computational resources used in this study included the following: the Darwin cluster at LANL, which is supported by the Computational Systems and Software Environment (CSSE) subprogram of the Advanced Simulation and Computing (ASC) program at LANL, which is funded by NNSA/DOE; resources provided by the LANL Institutional Computing program, which is funded by NNSA/DOE; and Northern Arizona University's Monsoon computer cluster, which is funded by Arizona’s Technology and Research Initiative Fund.

\section*{Author Contributions}
WSH and RGP designed the study. EDM and RS wrote the software. AI performed alpha testing. EDM and JC performed benchmarking. AH and HS upgraded libRoadRunner to enable integration into PyBioNetFit. EDM and WSH wrote the manuscript with input from the other authors. All authors read and approved the final manuscript. 

\section*{Conflict of Interest}
The authors declare no conflicts of interest.

\bibliography{references}
\end{document}